\documentclass[%
 reprint,
 amsmath,amssymb,
 aps,
]{revtex4-1}

\usepackage{xcolor,colortbl}
\definecolor{lightgray}{rgb}{0.85,0.85,0.85}
\newcommand{\tabitem}{~~\llap{\textbullet}~~}
\newcommand{\bigcell}[2]{\begin{tabular}{@{}#1@{}}#2\end{tabular}}
\usepackage{graphicx}
\usepackage{dcolumn}
\usepackage{bm}
\usepackage{hyperref}
\hypersetup{
    colorlinks=true,
    linkcolor=blue,
    filecolor=magenta,      
    urlcolor=blue,
}

\graphicspath{{Figures/}}

\begin{document}

\preprint{APS/123-QED}

\title{Emergence of Pseudo-Phononic Gaps in Periodically Architected Pendulums}

\author{H. Al Ba'ba'a}
\author{J. Callanan}
\author{M. Nouh}%
\email{Corresponding author: Mostafa Nouh, mnouh@buffalo.edu}
\affiliation{Department of Mechanical \& Aerospace Engineering, University at Buffalo (SUNY), Buffalo, NY 14260-4400}


\begin{abstract}
Rejection of unmitigated vibrational disturbances represents an ongoing dilemma in complex linkage systems. In this work, we present an inherent and self-reliant vibration isolation mechanism in architected periodic chains of serially pivoted pendulums. Absorption of external excitations is achieved by virtue of Bragg-type band gaps which stem from the emergent chain dynamics. Owing to its coupled dynamics, the self-repeating cell of a Phononic Crystal (PC) is not trivial and cannot be readily identified from the periodic arrangement by inspection. As such, this work entails the extraction of a ``pseudo" unit cell of an equivalent PC lattice from the derived motion equations of a finite chain, which departs from traditional wave dispersion methods. The model presented herein comprises a chain of linked pendulums with periodic variations of inertial/geometrical properties, carrying a payload at one end. We ultimately show evidence of forbidden wave propagation in prescribed frequency regimes, reminiscent of band gaps in PC lattices. The presented framework can be invaluable in applications that require vibration reduction in the delivery of payloads including gantry cranes, robotic arms and space tethers. 
\end{abstract}

\maketitle


\section{Introduction}

Undesirable vibrations and mechanical disturbances are an ever-lasting concern in the vast majority of engineering applications. The effect of vibrations on a dynamical system ranges from shortening its lifetime to sudden raptures, all of which impose serious operational challenges. The rational design of architected structures provides unique mechanisms by which unwanted arbitrary excitations can be rejected with minimal, and often non-existent, compromises in strength and resilience. Structural periodicity, whether in material composition \cite{ruzzene2000control}, topology \cite{bilal2011ultrawide}, boundary conditions \cite{albabaa2017PC}, placement of local resonances \cite{Pai2014} or via combinations of the previous \cite{liu2012wave}, enables a myriad of unprecedented wave dispersion capabilities; tailored to address challenges in vibroacoustic mitigation \cite{Hussein2014,bacquet2018metadamping}. Single and double negative materials \cite{huang2009negative, li2004double}, wave directivity \cite{Celli2015}, mechanical topological insulators \cite{pal2018amplitude}, and one-way elastic diodes \cite{trainiti2016non} are a few prime characteristics of engineered periodic systems. 

\begin{figure*}[]
\centering
\includegraphics[]{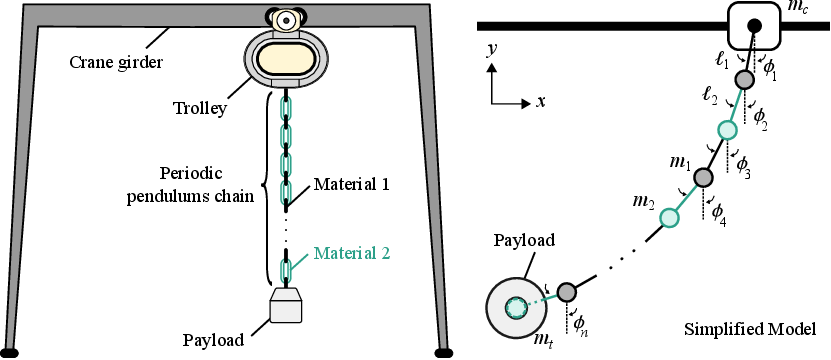}
\caption{A schematic of a gantry crane system with periodic pendulums chain with an end payload (left) and corresponding simplified pendulum chain model with a tip mass (right)}
\label{fig:PC_pendula}
\end{figure*}

This effort focuses on applications comprising mechanical pendulums and pendulum-like substructures suspended from a pivot and tasked with carrying and/or conveying a payload. Specifically, we address chains of coupled pendulums, the dynamics of which lie at the heart of several applications such as flexible robotic arms \cite{giorgio2018non}, unmanned aerial vehicles with payloads \cite{goodarzi2015geometric} and gantry (overhead) cranes \cite{masoud2017smooth}. In the latter, for instance, vibrations can be detrimental to the positioning accuracy of the terminal payload location. They elevate the conveyance duration of payloads and, consequently, create an operational bottleneck \cite{abdel2003dynamics}. The pendular motion of the payload has been by far regulated using active control techniques, exploiting tools such as command shaping\cite{masoud2017smooth} and feedback control \cite{Masoud2003}. In its most common form, payload is often hoisted via a rope or a linkage chain and is transported via a cart/trolley (see left panel of Figure~\ref{fig:PC_pendula}). The mechanism of the chain variant resembles that of a pendulum chain, where the tip of an individual pendulum (link) acts as a pivot to the subsequent one. Incorporating pendulums in novel metastructures has recently received increased attention. The most dominant example being 1D arrays of pendulums periodically coupled with springs which have been utilized to study a variety of intriguing phenomena pertaining to coupled nonlinear oscillators. These include solitons \cite{jallouli2017stabilization}, breathers \cite{russell1997moving}, energy transmission in band gaps \cite{geniet2002energy} and most recently helical edge states in topological insulators \cite{susstrunk2015observation}. To this end, however, exploiting the periodicity of pendulum chains as an inherent and self-reliant mechanism for vibration absorption remains uncharted territory. The aim of this work is to fill this gap by demonstrating phononic-like phenomena in the wave propagation profiles of such chains. Through a closed-form analysis of the dynamics of a periodic lumped pendulums chain with a payload (tip mass), as portrayed in Figure~\ref{fig:PC_pendula}, we ultimately show evidence of emergent phononic band gaps. Such gaps, resulting from Bragg-scattering and interference effects, restrict the propagation of waves within certain (and ultimately favorable) frequency regimes. 

Although the parameters of the chain of pendulums are periodic, the dynamics of each concentrated mass is strongly coupled with the whole chain as dictated by its motion equations \cite{masoud2017smooth}. As a result, the self-repeating unit cell of the chain arrangement (e.g. two masses $m_{1,2}$ and their links $\ell_{1,2}$ in a diatomic configuration as shown in Figure~\ref{fig:PC_pendula}) cannot be readily defined by the naked eye. As such, the most significant challenge unraveled here is the identification of a ``\textit{pseudo}" unit cell for Bloch-wave dispersion purposes and ultimately validating this cell choice from the actual pendulum response. Such identification process naturally departs from the conventional approach and instead relies on the dynamics of the finite
chain as a reference point. We start by deriving the equations of motion for a general chain of pendulums; the generalized model serves as the baseline for studying three distinct scenarios: (1) A uniform chain with and without a tip mass, (2) A chain with periodic variation of masses with a tip mass, and (3) A chain with periodic variation of the linkage lengths along with a tip mass. We mathematically show that the tip mass is crucial to obtaining a unit cell representation of an equivalent PC lattice which accurately captures the behavior of the chain under small vibrational amplitudes. The equivalent PC model is then compared with the actual linearized model of the periodic chain of pendulums in frequency and time domains. The latter is finally used to reconstruct the wave dispersion contours of the actual system, which validates the equivalent model for all three cases.

\section{Generalized Equations of Motion}
For a chain of pendulums with masses $m_i$ connected by rigid connections with length $\ell_i$, the position vector $\mathbf{r}$ for the $i^{th}$ mass is given by
\begin{equation}
\mathbf{r}_i =\Big(x - \sum_{j=1}^i  \ell_j \sin \phi_j \Big) \hat{\imath} - \sum_{j=1}^i \ell_j \cos \phi_j  \hat{\jmath}  
\end{equation}

and, hence, its velocity vector is

\begin{equation}
\mathbf{v}_i= \Big(\dot{x} -\sum_{j=1}^i \ell_j \dot{\phi}_j \cos \phi_j \Big) \hat{\imath} +  \sum_{j=1}^i \ell_j \dot{\phi}_j \sin \phi_j  \hat{\jmath}
\end{equation}

where $[\ \dot{} \ ] = \frac{d}{dt}$ is the time derivative, $\hat{\imath}$ and $\hat{\jmath}$ are unit vectors in the $x$ and $y$ directions, respectively, and the angles $\phi_j$ are labeled on the right panel of Figure~\ref{fig:PC_pendula}. Following which, the potential energy $\mathcal{V}$ of the pendulums chain can be expressed as

\begin{equation}
    \mathcal{V} = -g \sum_{i = 1}^n  \sum_{j = 1}^i m_i \ell_j  \cos\phi_j 
\end{equation}

while the kinetic energy, $\mathcal{T}$, is given by

\begin{widetext}
\begin{align}
\mathcal{T} = \frac{1}{2} \sum_{i = 1}^n m_i \mathbf{v}_i \cdot \mathbf{v}_i + \frac{1}{2} m_c \dot{x}^2 = \frac{1}{2} \sum_{i = 1}^n  \sum_{j = 1}^{i} m_i \ell_j \dot{\phi}_j \Bigg( \sum_{q = 1}^i \ell_q \dot{\phi}_q \cos(\phi_j-\phi_q) - 2 \dot{x} \cos(\phi_j)  \Bigg) +\frac{1}{2} \left(m_c + \sum_{i = 1}^n m_i \right) \dot{x}^2 
\end{align}
\end{widetext}
The Euler-Lagrange equation, i.e. $\frac{d}{dt} \Big(\frac{\partial \mathcal{L}}{\partial \dot{\phi}_i}\Big) -  \frac{\partial \mathcal{L}}{\partial \phi_i} = 0$, where $\mathcal{L}= \mathcal{T}-\mathcal{V}$ is the Lagrangian of the system, requires the evaluation of the partial derivative of the Lagrangian with respect to the angular position and velocity of the $i^{\text{th}}$ mass, i.e.
\begin{widetext}
\begin{equation}
    \frac{\partial \mathcal{L}}{\partial \dot{\phi}_i} = \sum_{r = i}^n \sum_{j = 1}^r  m_r \ell_i \left( \ell_j \dot{\phi}_j \cos(\phi_i - \phi_j) -  \dot{x} \cos(\phi_i) \right)
\end{equation}
\begin{equation}
    \frac{\partial \mathcal{L}}{\partial \phi_i} = \sum_{r = i}^n m_r \ell_i \left(\dot{\phi}_i \dot{x} - g \right) \sin(\phi_i) - \sum_{r = i}^n \sum_{j=1}^r m_r \ell_i \ell_j \dot{\phi}_i \dot{\phi}_j \sin(\phi_i - \phi_j)
\end{equation}
\begin{equation}
    \frac{d}{dt}\Bigg(\frac{\partial \mathcal{L}}{\partial \dot{\phi}_i}\Bigg) = \sum_{r = i}^n \sum_{j = 1}^r  m_r \ell_i \Bigg( \ell_j \Big[\ddot{\phi}_j \cos(\phi_i - \phi_j) - \dot{\phi}_j(\dot{\phi}_i - \dot{\phi}_j) \sin(\phi_i - \phi_j) \Big] + \dot{\phi}_i \dot{x} \sin(\phi_i) - \ddot{x} \cos(\phi_i)\Bigg)
\end{equation}
which leads to the equation of motion of the $i^{th}$ mass of the system, given by

\begin{equation}
\sum_{r = i}^n  \sum_{j = 1}^r m_r \Bigg(\ell_i \ell_j \Big[\ddot{\phi}_j \cos(\phi_i-\phi_j) + \dot{\phi}_j^2\sin(\phi_i-\phi_j) \Big]  + g \ell_i \sin(\phi_i) \Bigg) = \sum_{r = i}^n m_r \ell_i \ddot{x} \cos(\phi_i)
\label{eq:EOM_nonlinear_gen}
\end{equation}
\end{widetext}
Equation~(\ref{eq:EOM_nonlinear_gen}) serves as a generic equation of motion for the $i^{th}$ mass for any arbitrary set of parameters and number of masses in the pendulums chain. Assuming small oscillations, i.e. $\phi_i \approx 0$, Eq.~(\ref{eq:EOM_nonlinear_gen}) can be linearized and written as
\begin{equation}
\sum_{r = i}^n  \sum_{j = 1}^r m_r \left(\ell_i \ell_j \ddot{\phi}_j  + g \ell_i\phi_i \right) = \sum_{r = i}^n m_r \ell_i \ddot{x}
\label{eq:EOM_linearzied_gen}
\end{equation}

Using Eq.~(\ref{eq:EOM_linearzied_gen}) and setting the cart's acceleration, i.e. $\ddot{x}$, as an input, the equations of motion of the entire chain can be cast into a compact matrix form as
\begin{equation}
    \mathbf{M} \ddot{\boldsymbol{\phi}} + \mathbf{K} \boldsymbol{\phi} = \mathbf{f} \ddot{x}
    \label{eq:EOM_general}
\end{equation}
where 
\begin{subequations}
\begin{equation}
    \mathbf{M}_{ij} = \sum_{r = i}^n \sum_{j = 1}^r m_r \ell_i \ell_j
\end{equation}
\begin{equation}
\mathbf{K}_{ij} = 
\begin{cases}
g \sum_{r = i}^n m_r \ell_i & i=j\\
0 & i\neq j  \\
\end{cases}
\end{equation}
\begin{equation}
\mathbf{f}_{i} = \sum_{r = i}^n m_r \ell_i
\end{equation}
\begin{equation}
    \boldsymbol{\phi} =
    \begin{Bmatrix}
\phi_1 & \phi_2  & \cdots & \phi_n\\
\end{Bmatrix}^{\text{T}}
\end{equation}
\label{eq:mat_general}
\end{subequations}

\section{Dynamics of Periodic Pendulums}
\subsection{Case I: Uniform chain}

\begin{figure*}[]
\centering
\includegraphics[]{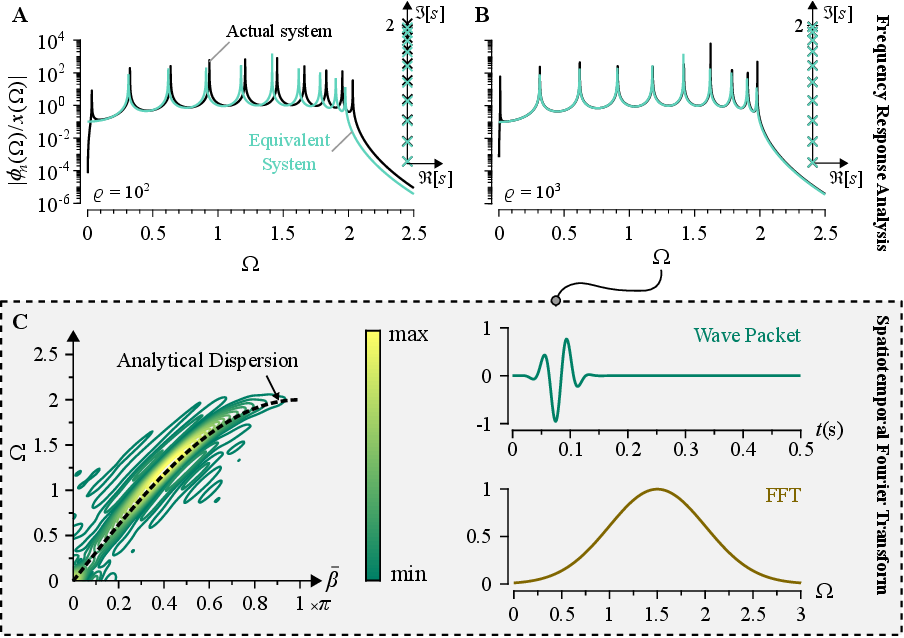}
\caption{Frequency response diagrams and corresponding poles distribution for (a) $\varrho = 10^2$ and (b) $\varrho = 10^3$ for a uniform chain of pendulums with $n=10$. (c) Numerical dispersion contours constructed from the actual chain of pendulums ($n=31$) showing agreement with the analytical dispersion branch (black dashed-line) obtained from the equivalent system via Eq.~(\ref{eq:disp_mono}). (Spectral and transient properties of the excitation are provided for reference)}
\label{fig:Case1}
\end{figure*}

For a uniform chain with identical rigid links lengths and lumped masses, i.e. $\ell_i = \ell$ and $m_i = m$, but without a tip mass, the linearized equation of motion of the $i^{\text{th}}$ mass given by Eq.~(\ref{eq:EOM_linearzied_gen}) reduces to
\begin{equation}
\sum_{j = 1}^n \bar{n}_{i,j} \ddot{\phi}_j + \frac{g}{\ell} \bar{n}_{i,i} \phi_i =  \frac{\ddot{x}}{\ell}\bar{n}_{i,i}
\end{equation}
where $\bar{n}_{i,j} = n-\max(i,j)+1$, which leads to the following system of equations:
\begin{equation}
    \mathbf{M} \ddot{\boldsymbol{\phi}} + \frac{g}{\ell}\mathbf{K} \boldsymbol{\phi} = \frac{\ddot{x}}{\ell} \mathbf{f} 
    \label{eq:uni_EOM_mat_no_tip}
\end{equation}
where 
\begin{subequations}
\begin{align}
\underset{n \times n}{\mathrm{\mathbf{M}}}=
\begin{bmatrix}
n & n-1  & n-2&\cdots & 1\\
n-1 & n-1 & n-2&\ddots& \vdots\\
n-2  & n-2 & n-2 &  \ddots & 1\\ 
\vdots & \ddots & \ddots &  \ddots & 1\\ 
1 &  \cdots  &1 & 1 &   1   \\
\end{bmatrix} 
\label{eq:M_uniform}
\end{align}
\begin{align}
\underset{n \times n}{\mathrm{\mathbf{K}}}= \mathbf{diag} 
    \begin{bmatrix}
n & n-1  & \cdots & 1
\end{bmatrix}
\label{eq:K_uniform}
\end{align}
\begin{equation}
    \mathbf{f} = 
    \begin{Bmatrix}
n & n-1  & \cdots & 1\\
\end{Bmatrix}^{\text{T}}
\end{equation}
\end{subequations}
The inverse of a matrix with the structure of $\mathbf{M}$ in Eq.~(\ref{eq:M_uniform}) has been shown to be a tridiagonal matrix \cite{da2001explicit,da2007eigenvalues}. For the uniform pendulums chain, the inverse of the matrix $\mathbf{M}$ in Eq.~(\ref{eq:M_uniform}) is derived as
\begin{align}
\underset{n \times n}{\mathrm{\mathbf{M}^{-1}}}=
\begin{bmatrix}
1 & -1 & 0 & \cdots & 0\\
-1 & 2 & \ddots& \ddots &\vdots\\
0 & \ddots & \ddots&\ddots &0\\
\vdots & \ddots & \ddots & 2 &  -1 \\
0 &  \cdots & 0 & -1 & 2 \\
\end{bmatrix} 
\label{eq:K}
\end{align}
Upon multiplying Eq.~(\ref{eq:uni_EOM_mat_no_tip}) by $\mathbf{M}^{-1}$, it simplifies to
\begin{equation}
    \ddot{\boldsymbol{\phi}} + \frac{g}{\ell} \mathbf{D} \boldsymbol{\phi} = \frac{1}{\ell} \mathbf{\hat{f}} \ddot{x}
\end{equation}
where 
\begin{align}
\underset{n \times n}{\mathrm{\mathbf{D}}}=
\begin{bmatrix}
n & 1-n  & 0 & \cdots &\cdots & 0\\
-n & 2(n-1) & 2-n& \ddots & &\vdots\\
0 & 1-n & \ddots  & \ddots & \ddots & \vdots\\
\vdots & \ddots & \ddots & \ddots & -2 & 0 \\ 
\vdots &  & \ddots & -3 & 4 & -1 \\
0 & \cdots & \cdots & 0 &-2 & 2 \\
\end{bmatrix} 
\label{eq:D}
\end{align}
and the forcing term reduces to $\mathbf{\hat{f}}^{\text{T}} =\begin{Bmatrix}
1 & 0 & \cdots & 0\\
\end{Bmatrix}$.
By inspecting Eq.~(\ref{eq:D}), it is evident that the structure of $\mathbf{D}$ is not periodic even though the physical chain has uniform properties. Consequently, a substructure consisting of one mass and its adjacent massless link can not be considered a self-repeating unit cell of the system for Bloch-wave analysis. Next, consider the same chain of masses with a tip mass $m_t$ at its end such that $m_t = \varrho m$. The linearized equation of motion of the $i^{\text{th}}$ mass now reads
\begin{equation}
\sum_{j = 1}^n (\varrho +\bar{n}_{i,j}) \ddot{\phi}_j + \frac{g}{\ell} (\varrho +\bar{n}_{i,i}) \phi_i =  \frac{\ddot{x}}{\ell}(\varrho +\bar{n}_{i,i})
\end{equation}


Casting the equations of motion into a matrix form similar to the previous analysis, it can be deduced that the inverse of the mass matrix in this case becomes
\begin{align}
\underset{n \times n}{\mathrm{\mathbf{M}^{-1}}}=
\begin{bmatrix}
1 & -1 & 0 & \cdots & 0\\
-1 & 2 & \ddots& \ddots &\vdots\\
0 & \ddots & \ddots&\ddots &0\\
\vdots & \ddots & \ddots & 2 &  -1 \\
0 &  \cdots & 0 & -1 & \frac{2+\varrho}{1+\varrho}\\
\end{bmatrix} 
\label{eq:Minv_uniform}
\end{align}
Now, if the payload is much heavier than the rest of the masses chain, i.e. $\varrho \ggg \bar{n}_{i,i}$, then the stiffness matrix can be approximated as $\mathbf{K} \approx \varrho \mathbf{I}$ with $\mathbf{I}$ being the unit matrix; the equations of motion now reads: 
\begin{equation}
    \ddot{\boldsymbol{\phi}} + \omega_0^2 \mathbf{M}^{-1} \boldsymbol{\phi} = \frac{1}{\ell} \mathbf{\hat{f}} \ddot{x}
    \label{eq:approx_EOM1}
\end{equation}
such that $\omega_0 = \sqrt{\frac{\varrho g}{ \ell}}$. Upon normalization using $\omega_0$, we arrive at the following reduced non-dimensional set of motion equations:
\begin{equation}
    \boldsymbol{\phi''} + \mathbf{M}^{-1} \boldsymbol{\phi} = \frac{1}{\ell} \mathbf{\hat{f}} x''
    \label{eq:approx_EOM}
\end{equation}
where $('')=\frac{d^2}{d\tau^2}$ and $\tau = \omega_0 t$. This system of equations identically resembles that of a monatomic lattice with free-free boundary conditions (since $\frac{2+\varrho}{1+\varrho} \rightarrow 1$ as $\varrho \rightarrow \infty$), and a repeating unit cell of an infinite medium can now be defined. The equivalent monatomic lattice comprises a chain of lumped masses $m$ connected via springs $k = \omega_0^2 m$. As such, the medium's dispersion relation can be derived as\cite{Hussein2014}
\begin{equation}
    \Omega =  \sqrt{2(1-\cos \bar{\beta})}
    \label{eq:disp_mono}
\end{equation}
where $\Omega = \frac{\omega}{\omega_0}$ and $\bar{\beta}$ are the dimensionless temporal and spatial frequencies, respectively. 

To verify these assumptions, we simulate the actual and equivalent systems in time and frequency domains. For the equivalent monatomic lattice, the end-to-end transfer function can be analytically derived as (see Al Ba'ba'a \textit{et al.}\cite{albabaa2018TDF} for details)
\begin{equation}
    \frac{\phi_n(s)}{x(s)} = \frac{s^2/\ell}{\prod_{i = 1}^{n} s^2 + 4\sin^2 \frac{\theta_i}{2}} =  \frac{1/\ell}{\prod_{i = 2}^{n} s^2 + 4\sin^2 \frac{\theta_i}{2}}
    \label{eq:TF_uniform}
\end{equation}
where $\theta_i$ depends on the boundary conditions of the system and is equal to $\frac{i-1}{n}\pi$ in this case \cite{yueh2005eigenvalues}. The roots of the denominator of Eq.~(\ref{eq:TF_uniform}) represent the system's poles (i.e. natural frequencies) and can be used to analytically determine the spectrum of the equivalent system. It can be seen that the rigid body mode at $s=0$ cancels perfectly with the $s^2$ term in the numerator; a result of the input being the acceleration of the cart, i.e. $\ddot{x}$. Figures~\ref{fig:Case1}a~and~\ref{fig:Case1}b show the frequency response for the system under consideration with $\varrho = 10^2$ and $\varrho = 10^3$ for both the equivalent and actual systems as well as the pole distribution on the complex $s$-domain. The results show excellent agreement between the equivalent and actual systems for $\varrho = 10^3$ while the frequency response and the pole distribution of the equivalent system slightly deviate from the actual one when $\varrho = 10^2$.

\begin{figure*}
\centering
\includegraphics[]{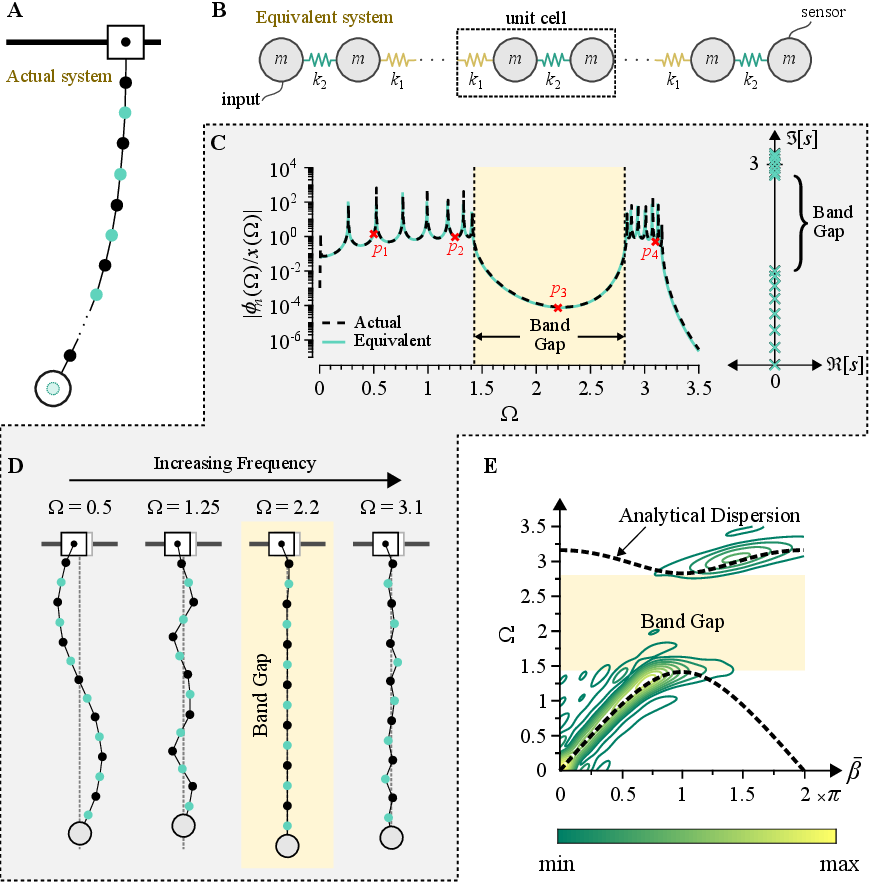}
\caption{Schematic of (a) the actual pendulums chain with varying masses and (b) the equivalent diatomic lattice with varying stiffnesses. (c) Frequency response diagram for the actual and equivalent systems for $n = 15$, $\mu = 0.25$, $\varrho = 10^3$ and $\ell = 1$ (left) and its corresponding poles distribution (right). (d) An illustrative schematic of the steady state responses of the pendulums with varying masses at four distinct excitation frequencies (marked $p_{1-4}$ on the frequency response). (e) Numerical dispersion contours constructed from the actual chain of pendulums showing agreement with the analytical dispersion relations (black dashed-line) obtained from the equivalent system via Eq.~(\ref{eq:disp_pc_case1})}
\label{fig:Case2}
\end{figure*}

To validate the dispersion analysis in Eq.~(\ref{eq:disp_mono}), on the other hand, the system is excited with a wide band excitation and its dispersion profile is constructed from the time-transient response of the system using the spatiotemporal Fourier transform \cite{airoldi2011design}. Figure~\ref{fig:Case1}c depicts the numerical dispersion contours for a chain of $n = 31$. The analytical dispersion relation of the equivalent system, i.e. Eq.~(\ref{eq:disp_mono}), is displayed as a dashed line for comparison and shows a very decent agreement.

\subsection{Case II: Periodic variation in masses \label{sec:mass_variation}}

The previous results for the uniform chain of pendulums can be extended to a chain with a periodic arrangement of masses, as portrayed in Figure~\ref{fig:Case2}a. Assume that the odd numbered masses are $m_1 = m$ while the even numbered are $m_2 = \mu m$, where $\mu=\frac{m_2}{m_1}$ is the mass ratio. Making use of Eq.~(\ref{eq:EOM_linearzied_gen}) and with few mathematical manipulations, the equations of motion simplify to Eq.~(\ref{eq:uni_EOM_mat_no_tip}) with the following new definitions for $\mathbf{M}$, $\mathbf{K}$ and $\mathbf{f}$:
\begin{widetext}
\begin{subequations}
\begin{align}
\mathbf{M}_{ij} = \varrho+(\mu+1)\Big  \lfloor \frac{n+1}{2}\Big \rfloor - \mu \left(\Big \lfloor \frac{\max(i,j)-1}{2}\Big \rfloor+\text{mod}(n,2)\right) - \Big \lfloor \frac{\max(i,j)}{2}\Big \rfloor
\label{eq:M_periodic_mass}
\end{align}
\begin{equation}
\mathbf{K}_{ij} = 
\begin{cases}
\varrho + (\mu+1)\Big  \lfloor \frac{n+1}{2}\Big \rfloor - \mu \left(\Big \lfloor \frac{i-1}{2}\Big \rfloor+\text{mod}(n,2)\right) - \Big \lfloor \frac{i}{2}\Big \rfloor & i=j\\
0 & i\neq j  \\
\end{cases}
\end{equation}
\begin{equation}
\mathbf{f}_{i} = \varrho + (\mu+1)\Big  \lfloor \frac{n+1}{2}\Big \rfloor - \mu \left(\Big \lfloor \frac{i-1}{2}\Big \rfloor+\text{mod}(n,2)\right) - \Big \lfloor \frac{i}{2}\Big \rfloor
\end{equation}
\label{eq:mat_periodic_masses}
\end{subequations}
\end{widetext}
where $\lfloor \cdot \rfloor$ is the floor function and ``$\text{mod}$" is the modulo operator. The mass matrix $\mathbf{M}$ in Eq.~(\ref{eq:M_periodic_mass}) has the following inverse (which is shown for an even $n$):
\begin{align}
\underset{n \times n}{\mathrm{\mathbf{M}^{-1}}}=
\begin{bmatrix}
1 & -1 & 0 & \cdots &\cdots & 0\\
-1 & 1+\frac{1}{\mu} & -\frac{1}{\mu} & \ddots & &\vdots\\
0 & -\frac{1}{\mu} & \ddots  & \ddots & \ddots & \vdots\\
\vdots & \ddots & \ddots & \ddots & -\frac{1}{\mu} & 0 \\ 
\vdots &  & \ddots & -\frac{1}{\mu} & 1+\frac{1}{\mu} & -1 \\
0 & \cdots & \cdots & 0 &-1 & \epsilon\\
\end{bmatrix} 
\label{eq:Minv_per_mass}
\end{align}

\noindent where $\epsilon =1+ \frac{1}{\mu+\varrho}$ for an even $n$ while for an odd $n$, we get $\epsilon = \frac{1}{\mu}+\frac{1}{1+\varrho}$. Note that $\epsilon$ reduces to either unity or $\frac{1}{\mu}$ for an even or odd $n$, respectively, as $\varrho \rightarrow \infty$ and $\mu = 1$ recovers $\mathbf{M}^{-1}$ for the uniform chain of pendulums, i.e. Eq.~(\ref{eq:Minv_uniform}). Since the payload is assumed to be much larger than the masses in the chain, the approximation $\mathbf{K} \approx \varrho \mathbf{I}$ remains intact leading to the equations of motion in Eq.~(\ref{eq:approx_EOM}) for the equivalent system with $\mathbf{M}^{-1}$ defined in Eq.~(\ref{eq:Minv_per_mass}). The resultant equations of motion resemble that of a typical diatomic lattice (see Figure~\ref{fig:Case2}b) with identical masses, alternating springs, and free-free boundary conditions \cite{albabaa2017PC}. The unit cell can be now defined as two identical lumped masses $m$ and two different springs $k_{1} = \frac{m}{\mu} \omega_0^2$ and $k_{2} = \mu k_1$, and, hence, the nondimensional dispersion relation for the equivalent system can be given by
\begin{equation}
    \Omega^4 - 2\Big[1+\frac{1}{\mu}\Big] \Omega^2 + \frac{4}{\mu} \sin^2 \Big(\frac{\bar{\beta}}{2}\Big) = 0
    \label{eq:disp_pc_case1}
\end{equation} 
with an emergent Bragg-type frequency band gap whose bounds are $\sqrt{2}$ and $\sqrt{2/\mu}$. A detailed derivation of the dispersion relation is provided in Appendix~\ref{App1}. It is very intriguing to observe that the variation of the masses in the periodic pendulums influences the equivalent stiffness of the system and, as a result, the mass ratio $\mu$ now dictates the ratio between the  stiffnesses in the equivalent system.

In line with the analysis performed for the uniform pendulums chain, the frequency and time responses of the equivalent and actual systems are computed to validate these predictions. Analytical transfer functions for a diatomic lattice have been derived in literature for any arbitrary combination of design parameters\cite{albabaa2017PC}. For brevity, the derivation details are omitted here and we only show the final forms of the end-to-end transfer functions. For an even $n$, we obtain
\begin{equation}
\frac{\phi_n(s)}{x(s)} =  \frac{\frac{1}{\ell}(\frac{1}{\mu})^{\frac{n}{2}-1}}{(s^2+2) \prod_{i=1}^{\frac{n}{2}-1} s^4 + 2(1+\frac{1}{\mu}) s^2 + \frac{4}{\mu} \sin^2 \big(\frac{\theta_i}{2} \big)}
\end{equation}
while for an odd number $n$, the transfer function reads
\begin{equation}
\frac{\phi_n(s)}{x(s)} =  \frac{\frac{1}{\ell}(\frac{1}{\mu})^{\lfloor \frac{n}{2} \rfloor}}{ \prod_{i=1}^{\lfloor \frac{n}{2} \rfloor} s^4 + 2(1+\frac{1}{\mu}) s^2 + \frac{4}{\mu} \sin^2 \big(\frac{\theta_i}{2} \big)}
\label{eq:FRF_PC_case1}
\end{equation}
where  $\theta_i = \frac{2i\pi}{n}$ in both cases. For $n = 15$, $\mu = 0.25$, $\varrho = 10^3$ and $\ell = 1$, Figure~\ref{fig:Case2}c shows the frequency responses of the equivalent and actual systems and their corresponding pole distributions. An excellent agreement between the two realizations is clearly observed. Unlike the case of the uniform pendulums, the natural frequency spectrum is split into two distinct groups of natural frequencies, before and after the band gap range, both of which have an essential role in the formation of the band gap itself as detailed in literature \cite{albabaa2017PC}. The band gap range in the frequency response spans the range $\sqrt{2}<\Omega<2\sqrt{2}$, which can be shown to match the band gap limits obtained from the dispersion relation in Eq.~(\ref{eq:disp_pc_case1}).

The steady state responses of the periodic pendulums chain at different frequencies are shown in Figure~\ref{fig:Case2}d. Within the band gap at $\Omega = 2.2$, the angular displacement corresponding to the payload at the tip of the chain is nearly zero, in addition to most of the chain masses prior to the payload, which is indicative of a band gap. On the other hand, the responses of the chain for $\Omega$ outside the range of the band gap clearly show a wave-like motion as expected from pass band frequencies. Finally, the dispersion behavior in Eq.~(\ref{eq:disp_pc_case1}) is tested by simulating the system's transient response to a wave packet in the time domain, as illustrated earlier. Figure~\ref{fig:Case2}e shows the numerical dispersion contours for the same parameters used to generate Figure~\ref{fig:Case2}c but with $n = 41$. As depicted by the figure, the entirety of the pendulums' dispersion energy coincides with dispersion bands predicted from Eq.~(\ref{eq:disp_pc_case1}) for the finite system.

\begin{figure*}
\centering
\includegraphics[]{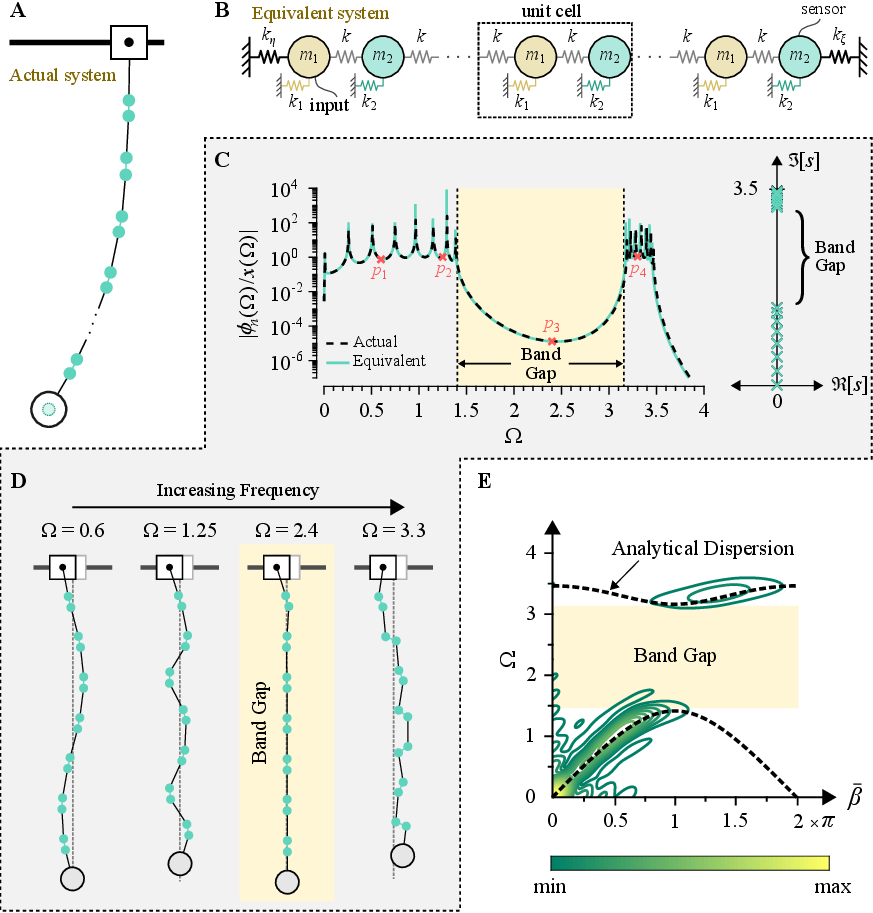}
\caption{Schematic of (a) the actual pendulums chain with linkage variations and (b) its equivalent PC lattice. (c) Frequency response diagram for the actual and equivalent systems for $n = 15$, $\kappa = 0.2$, $\varrho = 10^3$ and $\ell = 1$ (left) and its corresponding poles distribution (right). (d) An illustrative schematic of the steady state responses of the pendulums with varying linkage at four distinct excitation frequencies (marked $p_{1-4}$ on the frequency response). (e) Numerical dispersion contours constructed from the actual chain of pendulums showing agreement with the analytical dispersion relations (black dashed-line) obtained from the equivalent system via Eq.~(\ref{eq:disp_case2})}
\label{fig:Case3}
\end{figure*}

\begin{table*}[]
\centering
\caption{A summary of the considered periodic pendulums chain cases and their equivalent PC lattice with the presence of a payload mass. Boundary condition is abbreviated as B.C.}
\begin{tabular}{l l l l}
\hline\hline\
\hspace{0.3cm} \hspace{0.3cm} & \textbf{Actual Pendulum Chain} & \cellcolor{lightgray} \textbf{Dispersion-Equivalent Lattice} & \textbf{Additional Comments} \\
\hline 
\textbf{Case I} & Identical masses and linkages & \cellcolor{lightgray} Monatomic lattice & Free-free B.C. (finite structure) \\
\hline 
\textbf{Case II} &  \bigcell{l}{\textbf{Configuration:} \\ \tabitem Periodic masses $m_2/m_1 = \mu$ \\ \tabitem Uniform linkage $\ell_1 = \ell_2$} & 
\bigcell{l}{\cellcolor{lightgray} Diatomic lattice \\ \tabitem \cellcolor{lightgray} Varying stiffnesses with $k_2/k_1 = \mu$ \\ \tabitem \cellcolor{lightgray} Unity mass ratio $m_2/m_1 = 1$ \hfill }  & Free-free B.C. (finite structure) \\
\hline
\textbf{Case III} &  \bigcell{l}{\textbf{Configuration:} \\ \tabitem Periodic linkage $\ell_2/\ell_1 = \kappa$ \\ \tabitem Uniform masses with $m_1 = m_2$} & 
\bigcell{l}{\cellcolor{lightgray} Diatomic w/ elastic foundation \\ \cellcolor{lightgray} \tabitem Varying stiffness $k_2/k_1 = -1/\kappa$ \\ \cellcolor{lightgray} \tabitem Varying masses with $m_2/m_1 = 1/\kappa$\\ \cellcolor{lightgray} }  & \bigcell{l}{\tabitem
$k_1 = \frac{2m}{\kappa}\omega_0^2\left(\kappa-1\right)$ \\ \tabitem
$k_1 > 0$ for $\kappa > 1$ and $k_1 < 0$ for $\kappa < 1$
\\ 
\tabitem
$k_1k_2 < 0, \ \forall \kappa$
\\ 
\tabitem Fixed-fixed B.C. (finite structure)\\}\\
\hline\hline
\label{Tb:summary}
\end{tabular}
\end{table*}

\begin{figure*}
\centering
\includegraphics[]{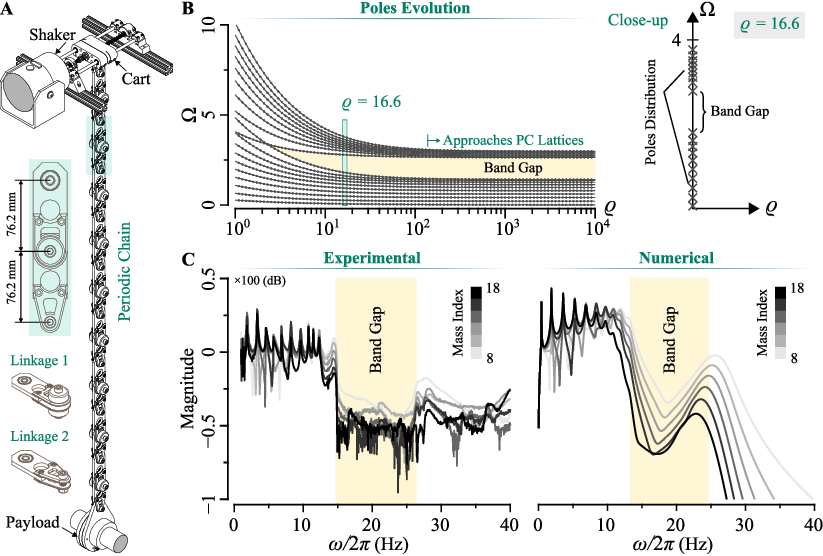}
\caption{(a) A schematic of the experimental setup comprising a shaker, cart and the pendulum chain with periodic mass (linkage) variations carrying a payload. (b) Evolution of poles as a function of the ratio $\varrho$ and a close-up of pole distribution at $\varrho=16.6$ (value used in the experimental testing). (c) Experimental and numerical transfer functions of the even-numbered masses starting at $i=8$ with respect to the cart for a frequency sweep up to $40$ Hz}
\label{fig:Exp1}
\end{figure*}

\subsection{Case III: Periodic variation in linkage lengths}

Consider the case of periodic pendulums with the lengths of odd and even numbered links being $\ell_1 = \ell$ and $\ell_2 = \kappa \ell$, respectively, where $\kappa$ is the length ratio of the periodic links; a visual illustration is provided in Figure~\ref{fig:Case3}a. As in the case of the periodically varying pendulums masses, the equations of motion simplify to Eq.~(\ref{eq:uni_EOM_mat_no_tip}) albeit with the following $\mathbf{M}$, $\mathbf{K}$ and $\mathbf{f}$ definitions:
\begin{subequations}
\begin{align}
\mathbf{M}_{ij} = 
\begin{cases}
\varrho + \min(\bar{n}_{i,i},\bar{n}_{j,j}) & \text{mod}(ij,2) = 1\\
\kappa (\varrho + \bar{n}_{i,j}) & \text{mod}(|i-j|,2) = 1 \\
\kappa^2 (\varrho + \bar{n}_{i,j}) & \text{mod}(|i-j|,2) = 0 \\
\end{cases}
\label{eq:M_periodic_link}
\end{align}
\begin{equation}
\mathbf{K}_{ij} = 
\begin{cases}
0 & i\neq j \\
\varrho + \bar{n}_{i,i} & \text{mod}(i,2) = 1\\
\kappa (\varrho + \bar{n}_{i,i})  & \text{mod}(i,2) = 0\\
\end{cases}
\label{eq:K_per_links}
\end{equation}
\begin{equation}
\mathbf{f}_{i} =
\begin{cases}
\varrho + \bar{n}_{i,i} & \text{mod}(i,2) = 1\\
\kappa (\varrho + \bar{n}_{i,i}) & \text{mod}(i,2) = 0\\
\end{cases}
\end{equation}
\label{eq:mat_periodic_masses2}
\end{subequations}
and the inverse mass matrix in this case (shown for an even number $n$) is:
\begin{align}
\underset{n \times n}{\mathrm{\mathbf{M}^{-1}}}=
\begin{bmatrix}
1 & -\frac{1}{\kappa} & 0 & \cdots &\cdots & 0\\
 -\frac{1}{\kappa} &  \frac{2}{\kappa^2} & -\frac{1}{\kappa}& \ddots & &\vdots\\
0 & -\frac{1}{\kappa} & 2 & \ddots & \ddots & \vdots\\
\vdots & \ddots & \ddots & \ddots & -\frac{1}{\kappa} & 0 \\ 
\vdots &  & \ddots & -\frac{1}{\kappa} & 2 & -\frac{1}{\kappa} \\
0 & \cdots & \cdots & 0 &-\frac{1}{\kappa} & \varepsilon\\
\end{bmatrix} 
\label{eq:Min_length_var}
\end{align}

For an even $n$, $\varepsilon = \frac{1}{\kappa^2}(1+\frac{1}{1+\varrho})$ while for an odd one, $\varepsilon = 1+\frac{1}{1+\varrho}$. Unlike the case of periodic masses, the stiffness matrix in Eq.~(\ref{eq:K_per_links}) is approximated by a periodic diagonal matrix which is valid for a large $\varrho$ value and, for an even $n$, is given by
\begin{align}
\underset{n \times n}{\mathrm{\mathbf{K}}}= \varrho 
\begin{bmatrix}
1 & 0 & \cdots & \cdots & 0\\
0 & \kappa & \ddots & & \vdots\\
\vdots & \ddots & \ddots & \ddots & \vdots \\
\vdots & & \ddots & 1 & 0 \\
0 & \cdots & \cdots & 0 & \kappa \\
\end{bmatrix} 
\label{eq:K_case2}
\end{align} 

Since $\mathbf{M}^{-1}$ acts as the stiffness matrix of an equivalent PC lattice, the inverse of $\mathbf{K}$ in Eq.~(\ref{eq:K_case2}) can be treated as the equivalent mass matrix, and will ultimately yield the unit cell of the equivalent PC lattice. Observing the pattern of $\mathbf{M}^{-1}$ in Eq.~(\ref{eq:Min_length_var}) and the inverse of $\mathbf{K}$ in Eq.~(\ref{eq:K_case2}), it can be concluded that these matrices resemble the stiffness and mass matrices, respectively, for a diatomic lattice with periodic elastic foundations and varying masses, and with two boundary springs $k_\eta$ and $k_\xi$. For visualization, this arrangement is depicted in Figure~\ref{fig:Case3}b. Based on Eq.~(\ref{eq:Min_length_var}), the PC chain has an equivalent uniform springs of $k = \frac{m}{\kappa}\omega_0^2$ and the stiffness of the periodic elastic foundations are $k_1 = 2k\left(\kappa-1\right)$ and $k_2 = \frac{-k_1}{\kappa}$, which implies that either $k_1$ or $k_2$ has to be negative depending on the value of $\kappa$. The latter is interesting since a negative stiffness is usually induced by active elements, e.g. piezoelectric shunts \cite{chen2014piezo}. In the present case, however, the negative stiffness is a direct consequence of the unique dynamics of the pendulums chain. Similarly, it can be shown that the spring constant $k_\eta = -\frac{k_1}{2}$ for all $n$, while $k_\xi = k_\eta$ and $k_\xi = -\frac{k_2}{2}$ for odd and even values of $n$, respectively. In addition to the periodic elastic foundation, the variation in the length of the links creates a periodic variation in the masses for the equivalent lattice with values $m_1 = m$ and $m_2 = m_1/\kappa$. For these values of springs and masses, and for the unit cell marked on the schematic given by Figure~\ref{fig:Case3}b, the following dispersion relation can be derived:
\begin{equation}
    \Omega^4 - 2\Big[1+\frac{1}{\kappa}\Big] \Omega^2 + \frac{4}{\kappa} \sin^2 \Big(\frac{\bar{\beta}}{2}\Big) = 0
    \label{eq:disp_case2}
\end{equation}
and the corresponding band gap bounds are $\sqrt{2}$ and $\sqrt{2/\kappa}$. Further details pertaining to this derivation are provided in Appendix~\ref{App2}.

Given the boundary conditions of this PC model, a closed-form analytical solution for the system's poles remains elusive. As a result, the frequency response in this scenario is derived in semi-analytical form since the poles of the system are calculated numerically. Once the poles are computed, the following equation depicts the system's transfer function for all $n$ (See Miu \cite{miu1993mechatronics} for more details):
\begin{equation}
\frac{\phi_n(s)}{x(s)} =  \frac{s^2(\frac{1}{\kappa})^{\lfloor \frac{n}{2} \rfloor}}{\ell\prod_{i=1}^{n} s^2+\Omega_i^2}
\label{eq:case3}
\end{equation}

\begin{figure*}
\centering
\includegraphics[width=0.7\textwidth]{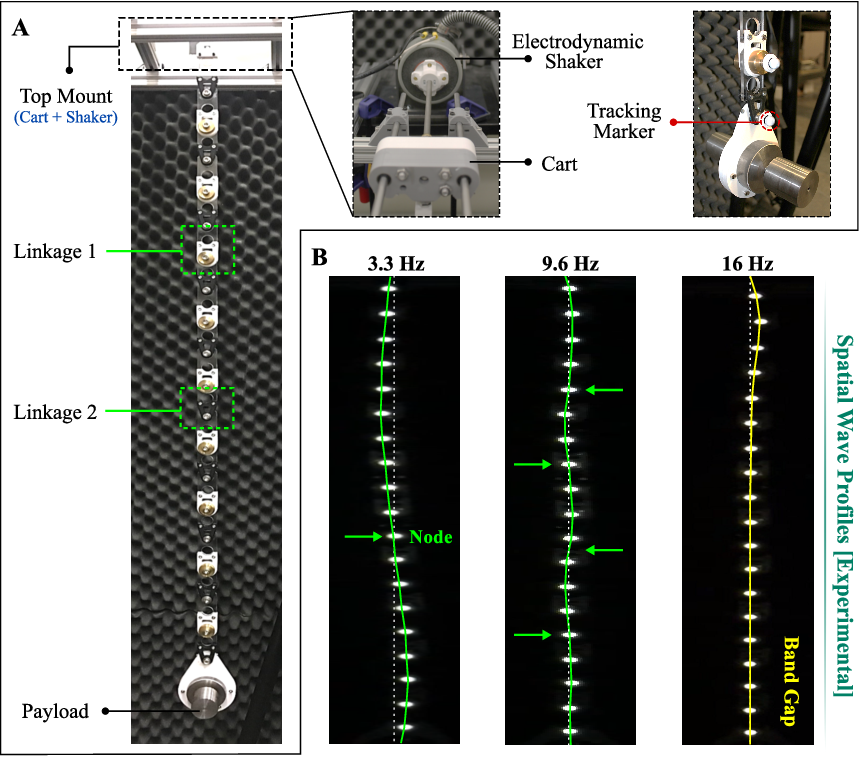}
\caption{(a) Front view of the fully assembled pendulum chain (left). Close-ups of the mounting system, electrodynamic shaker, cart as well as the linkages with the attached reflective markers (right). (b) Experimental snapshots of the spatial wave profiles outside ($3.3$ and $9.6$ Hz) and inside ($16$ Hz) a band gap}
\label{fig:Exp2}
\end{figure*}

In Eq.~(\ref{eq:case3}), $\Omega_i$ is a normalized pole of the system. For $\kappa = 1/5$, $n=15$, $\ell = 1$ and $\varrho = 10^3$, Figure~\ref{fig:Case3}c shows the end-to-end frequency response function for the actual and equivalent systems, where a very good match is observed. As expected, a band gap is created in the range $\sqrt{2}<\Omega<\sqrt{10}$ which coincides with the large attenuation region in the frequency response diagram and matches the range predicted from the dispersion relation in Eq.~(\ref{eq:disp_case2}). The steady state responses provide a further confirmation of the occurrence of the band gap as can be seen from the displacement fields at frequencies inside (i.e. $\Omega = 2.4$) and outside the band gap (i.e. $\Omega = 0.6, 1.25 \ \text{and} \ 3.3$) as shown Figure~\ref{fig:Case3}d. It is worth noting that although a zero frequency band gap is traditionally onset in PC lattices with elastic foundations \cite{albabaa2017PC}, the negative stiffness of the elastic foundation in this scenario neutralizes the effect of the grounded springs and effectively renders it a foundation-free lattice. Finally, the numerical dispersion contours for the same system parameters used in Figure~\ref{fig:Case3}c and for $n = 41$ is depicted in Figure~\ref{fig:Case3}e. As anticipated, the contours are again concentrated and centered around the dispersion bands predicted from Eq.~(\ref{eq:disp_case2}). On a final note, a succinct presentation of the three presented cases is provided in Table~\ref{Tb:summary}. The latter summarizes the dual correlations between the actual pendulum chain and the hypothetical dispersion-equivalent system for cases I, II and III.

\section{Experimental Validation} 

To verify the emergence of the aforementioned class of Bragg band gaps in a periodically architected pendulum chain, a full-scale experimental apparatus has been designed and constructed. The experimental setup reflects a periodic arrangement of masses, similar to Case II in Section~\ref{sec:mass_variation}. The chain, shown in Figure~\ref{fig:Exp1}a, comprises nineteen masses alternating between heavy (Linkage 1) and light (Linkage 2) linkages with a mass ratio $\mu = 1/3.46$. The linkages were constructed from quarter-inch polycarbonate plates and a set of 3D-printed PLA plastic sections. The lumped-masses consist of one bearing, machine screw fasteners, and --in the case of Linkage 1-- brass cylinders to increase the mass. The periodic pendulum chain was suspended from a cart, such that its motion is constrained in one direction and actuated by an electrodynamic shaker. A payload was fixed to the final mass in the chain such that the ratio $\varrho$ is equal to $16.6$. A complete to-scale drawing of the experimental apparatus is shown in Figure~\ref{fig:Exp1}a. 

The ratio $\varrho = 16.6$ used in this experiment is sufficiently large to initiate a band gap in the periodic chain, yet smaller (in width) in comparison to the fully developed band gap at large values of $\varrho$. As such, the evolution of the system poles as the ratio $\varrho$ approaches infinity is crucial in the experimental design as well as quantifying the anticipated band limits associated with different system parameters. Figure~\ref{fig:Exp1}b illustrates such evolution by capturing the variation in the distribution of natural frequencies as a function of $\varrho$. As expected, the finite system approaches the analytical solution in Eq.~(\ref{eq:FRF_PC_case1}) as the payload mass increases. The numerically computed band gap at $\varrho = 16.6$ is also seen in the close-up shown in Figure~\ref{fig:Exp1}b.

In order to experimentally capture and measure the band gap, a frequency sweep excitation was carried out and the transfer functions between the cart and various linkages throughout the chain were measured. The electrodynamic shaker was actuated over a range of frequencies ranging from 1 to 40 Hz via a signal generator and an accompanying amplifier system. Two piezoelectric accelerometers were fixed to the chain, one attached to the cart and the second to the even-numbered masses, starting at the eighth mass up to the chain's end. The results of the experimental frequency response are given in Figure~\ref{fig:Exp1}c. The left plot of Figure~\ref{fig:Exp1}c shows the magnitude of the experimentally obtained transfer function as a function of frequency, whereas the right plot shows simulation results for the system described in Section 3.2 (Case II), albeit with a moderate degree of damping. The experimental results agree reasonably well with the numerical simulations. Specifically, the emergent band gap approximately spans the ranges $14.7$--$26.5$ Hz and $13.1$--$24.5$ Hz in the experimental and numerical cases, respectively. 

Finally, a visualization of the band gap is obtained via a video recording of the chain's response in real time. Figure~\ref{fig:Exp2}a shows a front view of the fully assembled setup, as well as close-ups of its top mount (i.e. shaker and cart system) and the different linkages. A set of reflective markers were attached to the center of the lumped masses to capture the spatial wave profiles under different excitation frequencies. A concentrated light beam is shined on the reflective markers with the surrounding space being dimmed to clearly reflect the oscillations of the chain, and the response of the chain is tracked in slow motion via a fixed camera facing the setup. Snapshots of the recordings are shown in Figure~\ref{fig:Exp2}b for $3.3$ Hz, $9.6$ Hz and $16$ Hz excitations. The first two cases represent a couple of vibration modes where the injected wave propagates freely to the payload at the bottom end of the chain. The last case, i.e. $16$ Hz, corresponds to a band gap frequency and the blocked wave propagation can be clearly observed. A full video demonstration of the entire experiment can be found \href{https://www.frontiersin.org/articles/10.3389/fmats.2019.00119/full#supplementary-material}{here}.

\section{Concluding Remarks}
In this paper, the ability of a passive and non-dissipative periodic pendulum chain to exhibit self-induced vibration isolation capabilities was demonstrated. Given the non-trivial and coupled nature of the pendulum chain dynamics, an obvious definition of a unit cell --typically needed to conduct and predict Bragg band gaps in self-repeating structures-- does not exist. Instead, a pseudo unit cell of an equivalent PC lattice was identified and extracted from the finite chain's response in a novel and innovative manner. The analysis showed that the payload carried by the pendulum chain becomes key to the formation of the equivalent PC system and, in its presence, the system shows a similar dispersive behavior. As a result, periodic variations in the pendulums masses or links (with a tip mass) create a frequency band gap which has been verified in both frequency and time domains. As a very intriguing observation, we showed that: (1) The variation in masses resembles the effect of varying the stiffness in a conventional PC. (2) On the other hand, a periodic change in the connecting links becomes reminiscent of a PC lattice with a periodic elastic foundation as well as a periodic variation in the masses. Both cases, however, act identically since the hypothetical elastic foundation comprises a negative stiffness which nullifies its own effect. This is particularly interesting given that negative stiffness systems are typically the product of auxetic structures or actively controlled elements, as opposed to the current case where it emerges solely from the pendulum's unique arrangement. 

The unconventional dynamics of the pendulum chain have been experimentally demonstrated via transfer function measurements, showing a clear evidence of the predicted band gaps, in addition to snapshots of the spatial profiles both inside and outside the band gap which support the derived transfer function. The potential of the proposed pendulum chains can extend beyond industrial crane applications to ones which involve vibration control of payload deliveries including, but not limited to, robotic arms (\cite{o2007wave}), descending payloads of unmanned aerial vehicles (\cite{goodarzi2015geometric}) as well as tethered space elevators (\cite{kim1995modeling,williams2009dynamic}).

\section*{Acknowledgements}
The authors acknowledge the support of this work from the US National Science Foundation through awards no.~1647744 and 1847254 (CAREER).

\bibliography{sample}
\vspace{0.5 cm}

\appendix

\section{Pendulums with Periodic Masses}
\label{App1}
\noindent The unit cell equations of motion corresponding to the system shown in Figure~3b in the manuscript can be written as
\begin{align}
    & m \ddot{u}_i + (k_1 + k_2) u_i - k_1 v_{i-1} - k_2 v_{i} = 0\\
    & m \ddot{v}_i + (k_1 + k_2) v_i - k_1 u_{i+1} - k_2 u_{i} = 0
\end{align}
where $u_i$ and $v_i$ represent the displacement of the odd and even numbered masses, respectively. Applying the Bloch-wave solution and assuming harmonic motion result in an eigenvalue problem, the determinant of which yields the dispersion relation:
\begin{equation}
m^2 \omega^4 - 2m(k_1+k_2)\omega^2 + 4 k_1 k_2 \sin^2 \Big(\frac{\bar{\beta}}{2}\Big) = 0
\end{equation}
For the case of $k_{1} = \frac{\omega_0^2 m}{\mu}$ and $k_{2} =\mu k_1$ and dividing by $m^2$ throughout yields

\begin{equation}
    \omega^4 - 2\omega_0^2\Big[1+\frac{1}{\mu}\Big] \omega^2 + \frac{4}{\mu} \omega_0^4 \sin^2 \Big(\frac{\bar{\beta}}{2}\Big) = 0
    \label{eq:disp_case1_dim}
\end{equation}
Normalizing Eq.~(\ref{eq:disp_case1_dim}) using $\omega_0$, we obtain the nondimensional dispersion relation in Eq.~(26) in the manuscript.

\section{Pendulums with Periodic Links}
\label{App2}

\noindent For the unit cell marked in Figure~4b in the manuscript, the equations of motion can be written as
\begin{align}
    & m_1 \ddot{u}_i + (2k + k_1) u_i - k v_{i-1} - k v_{i} = 0\\
    & m_2 \ddot{v}_i + (2 k + k_2) v_i - k u_{i+1} - k u_{i} = 0
\end{align}
where $u_i$ and $v_i$ are as defined in Section \ref{App1}. The dispersion relation can be derived as
\begin{widetext}
\begin{equation}
m_1 m_2 \omega^4 - \big[ m_1(2k+k_2)+ m_2(2k+k_1)\big]\omega^2 + 2k(k_1 + k_2) + k_1 k_2 + 4 k^2 \sin^2 \Big(\frac{\bar{\beta}}{2}\Big) = 0
\label{eq:disp_case2_dim}
\end{equation}
\end{widetext}
Given the matrix definition in Eq.~(29) in the manuscript, $(2k + k_1)$ is equal to $2 \omega_0^2 m$ based on the odd numbered diagonal elements of the matrix. Knowing that $k = \frac{m}{\kappa} \omega_0^2$ from the off-diagonal elements, we solve for $k_1$ which gives $k_1 = 2 k \left(\kappa-1\right) $. Similarly, $(2k + k_2)$ is equal to $\frac{2}{\kappa^2}\omega_0^2 m  $ based on the even numbered diagonal elements, which results in $k_2 = 2 \frac{m}{\kappa}\omega_0^2 (\frac{1}{\kappa}-1) = \frac{-k_1}{\kappa} $ when solving for $k_2$. The inverse of the matrix $\mathbf{K}$ in Eq.~(30) in the manuscript can be treated as the equivalent mass matrix of the PC lattice which gives $m_1 = m$ and $m_2 = \frac{m}{\kappa}$. Substituting the values of $m_1$, $m_2$, $k_{1}$ and $k_{2}$ and dividing by $\frac{m^2}{\kappa}$, the dispersion relation reduces to
\begin{equation}
    \omega^4 - 2\omega_0^2\Big[1+\frac{1}{\kappa}\Big] \omega^2 + \frac{4}{\kappa} \omega_0^4 \sin^2 \Big(\frac{\bar{\beta}}{2}\Big) = 0
    \label{eq:disp_case2_dim2}
\end{equation}
Upon normalization, we obtain the dispersion relation in Eq.~(32) in the manuscript. The terms $k_1 k_2$ and $2k(k_1 + k_2)$ in Eq.~(\ref{eq:disp_case2_dim}), which are essential for creating a zero frequency band gap, cancel out each other for the given values of $k_1$ and $k_2$. The final form of Eq.~(\ref{eq:disp_case2_dim2}) is very similar to Eq.~(\ref{eq:disp_case1_dim}) even though the equations of motion for the finite structures are completely different.

\end{document}